\documentclass{aa}

\usepackage{multirow}
\usepackage{graphicx}
\usepackage[varg]{txfonts}
\usepackage{natbib}

\def\im{\mathrm{i}}

\def\expo#1{\mathbf{e}^{#1}}

\def\H{\mathcal{H}}
\def\G{\mathcal{G}}

\def\C22i{C_{22}}
\def\zetai{\zeta}
\def\kf{k_f}
\graphicspath{{./figures/}}

\newcommand\figI{
  \begin{figure*}
    \centering
    \includegraphics[width=\linewidth]{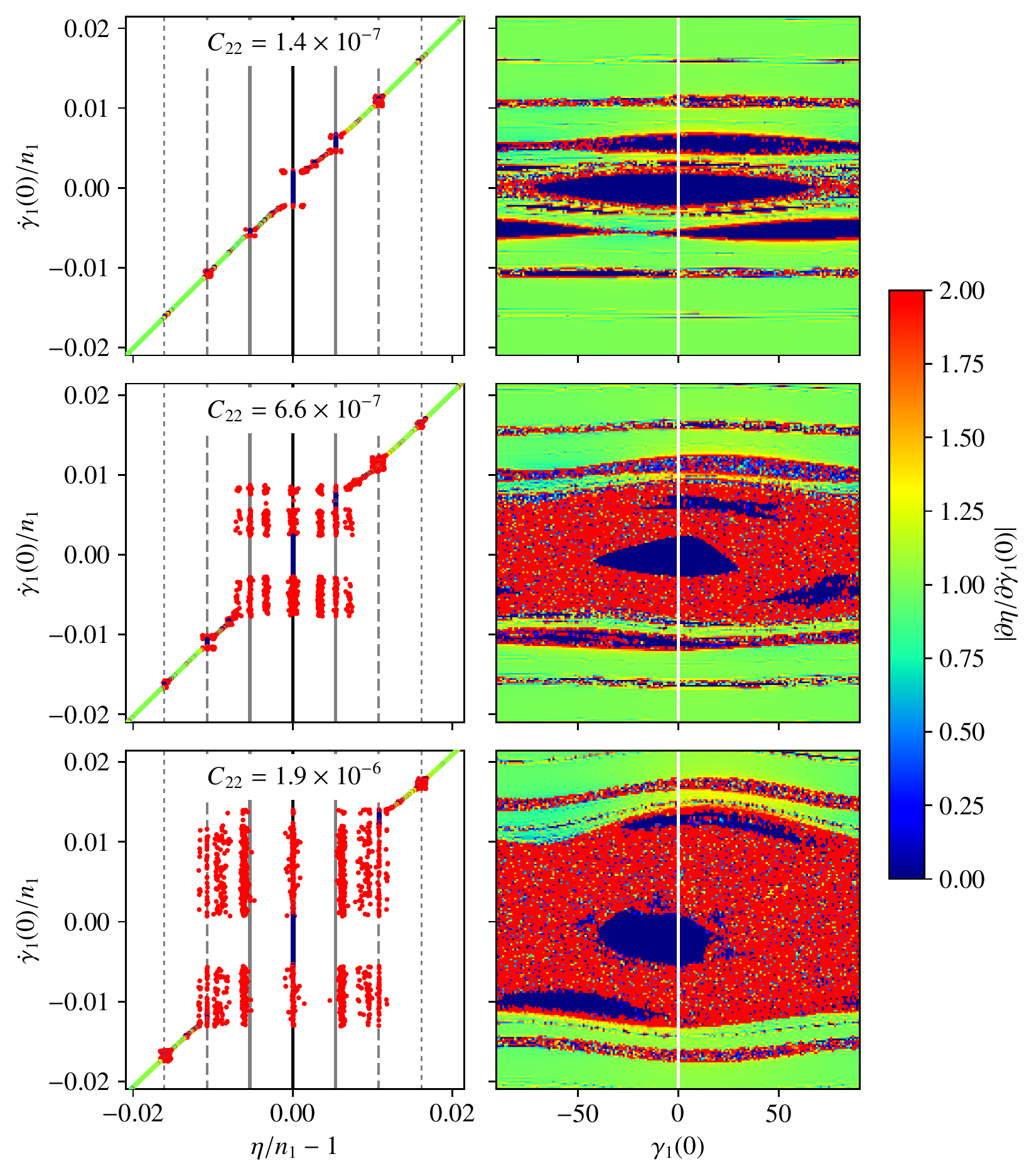}
    \caption{Spin dynamics of \object{KOI-227}\,$b$ in the conservative case,
      with $C_{22}=1.4\times 10^{-7}$
      (\textit{top}, permanent deformation),
      $6.6\times 10^{-7}$
      (\textit{middle}, maximum deformation in the sub/super-synchronous resonances),
      and $1.9\times 10^{-6}$
      (\textit{bottom}, maximum deformation in the synchronous resonance).
      The left column shows the main frequency $\eta$ of $\theta(t)$
      for different initial values of $\dot{\gamma}(0)$ (and with $\gamma(0) = 0$).
      The colour gives the derivative $\partial\eta/\partial\dot{\gamma}(0)$.
      Blue dots correspond to resonant motion, green dots to non-resonant regular motion,
      and red dots to chaotic motion.
      The vertical black line corresponds to the synchronization ($\eta=n$).
      The two grey lines correspond to
      the main sub/super-synchronous resonances ($\eta = n \pm \nu/2$).
      The dashed and dotted grey lines correspond to
      higher order sub/super-synchronous resonances ($\eta = n + k \nu/2$, $k=\pm2$, $\pm3$).
      The right column shows the same colour index ($\partial\eta/\partial\dot{\gamma}(0)$),
      but both $\gamma(0)$ and $\dot{\gamma}(0)$ are varied (2d maps).
      The vertical white line highlights the initial conditions taken
      in the left column ($\gamma(0) = 0$).
      }
    \label{fig:I}
  \end{figure*}
}

\newcommand\figII{
  \begin{figure*}
    \centering
    \includegraphics[width=\linewidth]{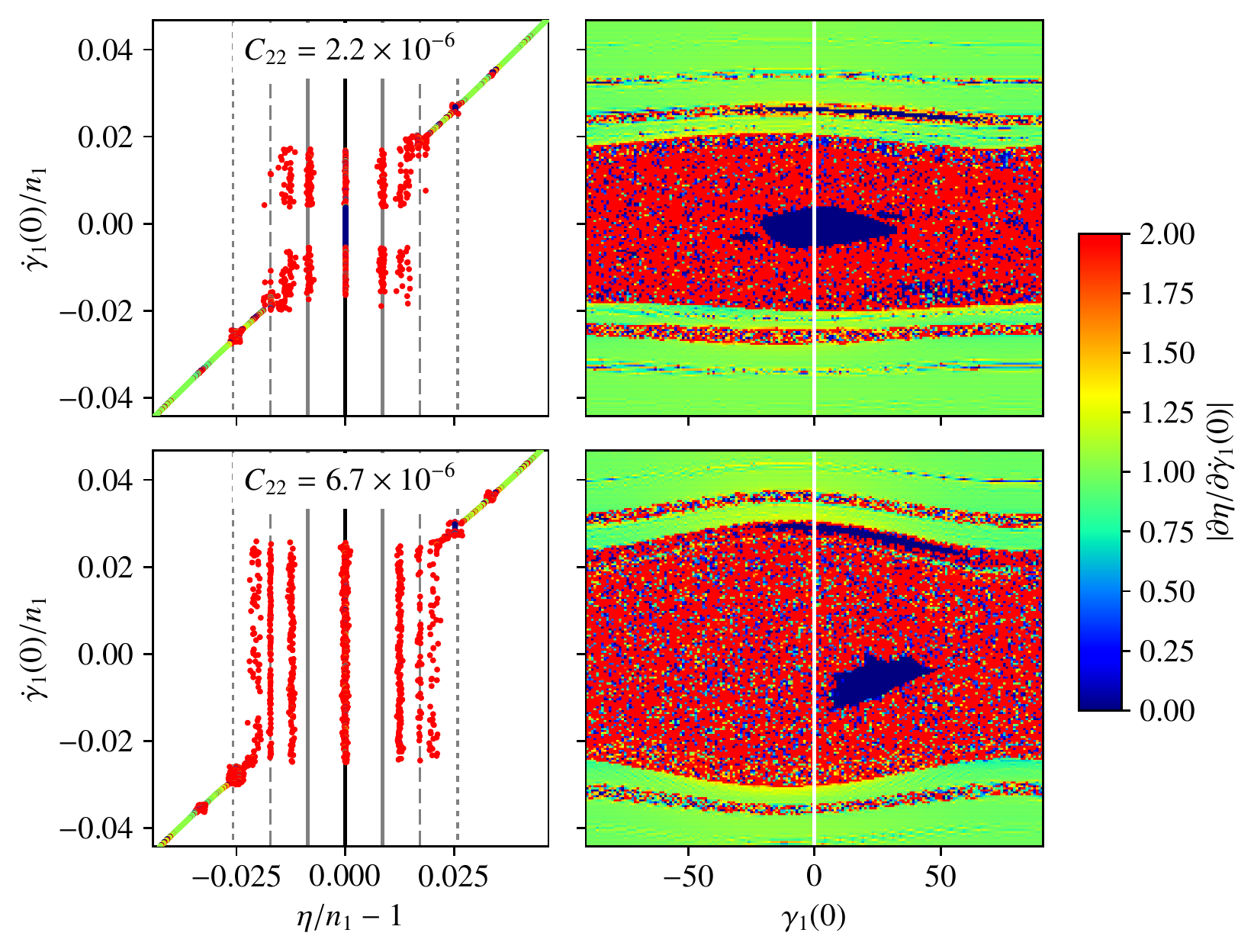}
    \caption{Same as Fig.~\ref{fig:I} but for \object{Kepler-88}\,$b$.
    We neglect the permanent deformation of the planet ($C_{22,r}=0$),
    and show the spin dynamics for
    $C_{22} = 2.2\times 10^{-6}$
    (\textit{top}, maximum deformation in the sub/super-synchronous resonances),
    and $6.7\times 10^{-6}$
    (\textit{bottom}, maximum deformation in the synchronous resonance).}
    \label{fig:II}
  \end{figure*}
}

\newcommand\figIII{
  \begin{figure*}
    \centering
    \includegraphics[width=\linewidth]{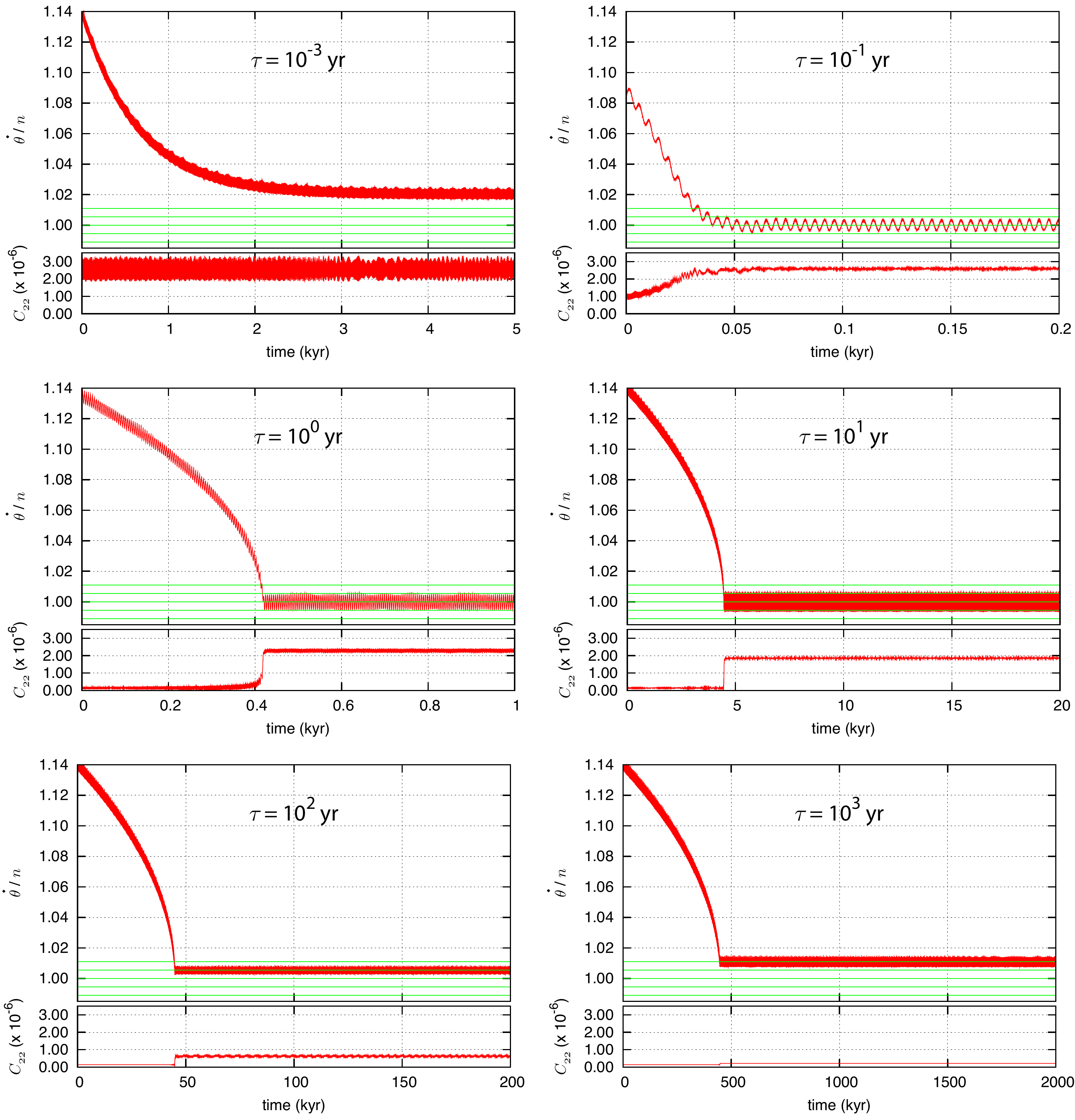}
    \caption{Examples of the spin evolution of \object{KOI-227}\,$b$ and corresponding global $C_{22}$ value for different $\tau$ values. We use the initial conditions from Table~\ref{tab:II}, and the initial rotation period is set at 15.5~d. The green lines give the position of the spin-orbit resonances $n$, $n\pm\nu/2$, and $n\pm\nu$.
    }
    \label{fig:III}
  \end{figure*}
}

\newcommand\figIV{
  \begin{figure*}
    \centering
    \includegraphics[width=\linewidth]{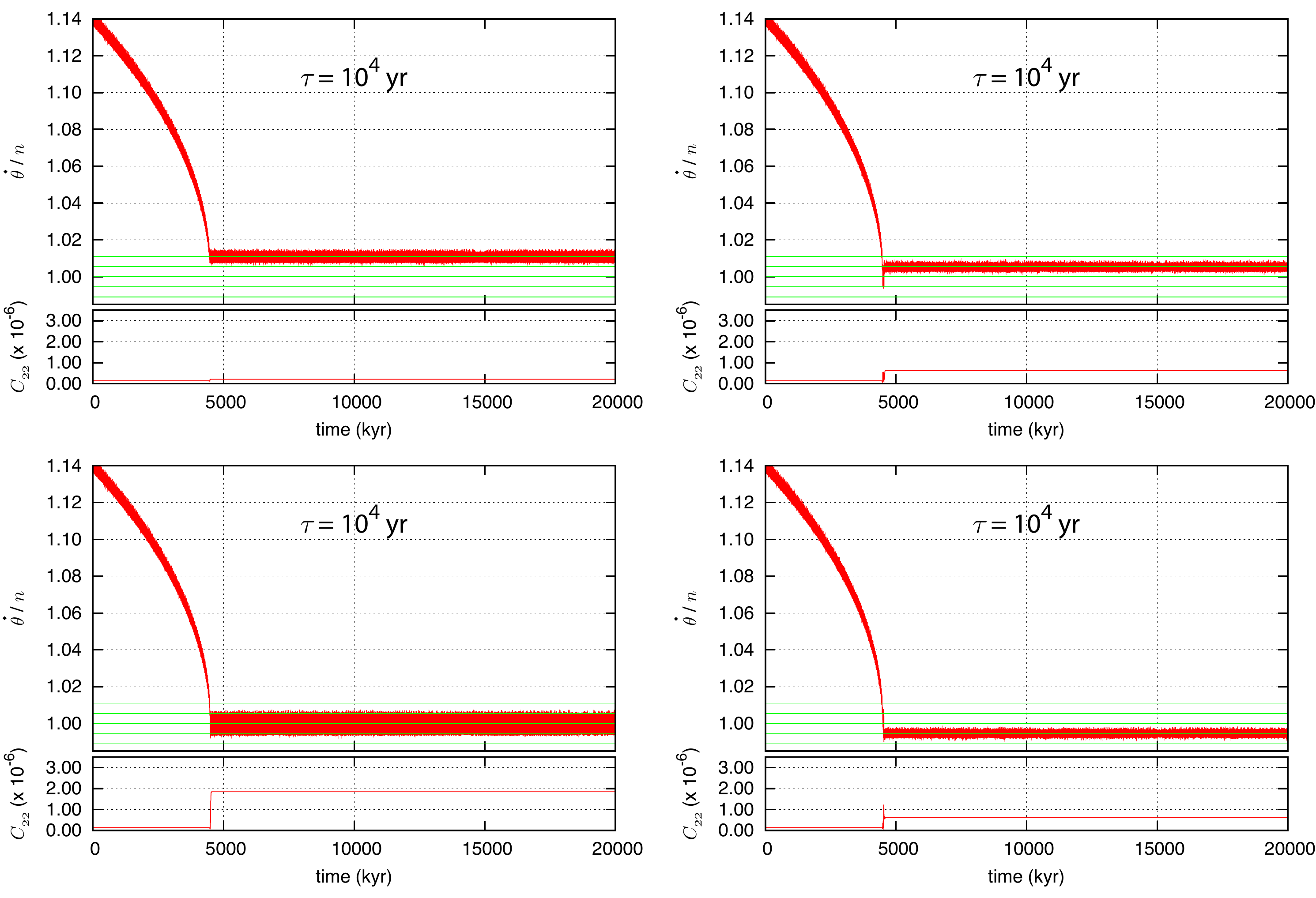}
    \caption{Different final spin evolution of \object{KOI-227}\,$b$ and corresponding global $C_{22}$ value for $\tau=10^4$~yr. We use the initial conditions from Table~\ref{tab:II}, and the initial rotation period is set at 15.5~d. The green lines give the position of the spin-orbit resonances $n$, $n\pm\nu/2$, and $n\pm\nu$.
    }
    \label{fig:IV}
  \end{figure*}
}

\newcommand\figV{
  \begin{figure}
    \centering
    \includegraphics[width=\linewidth]{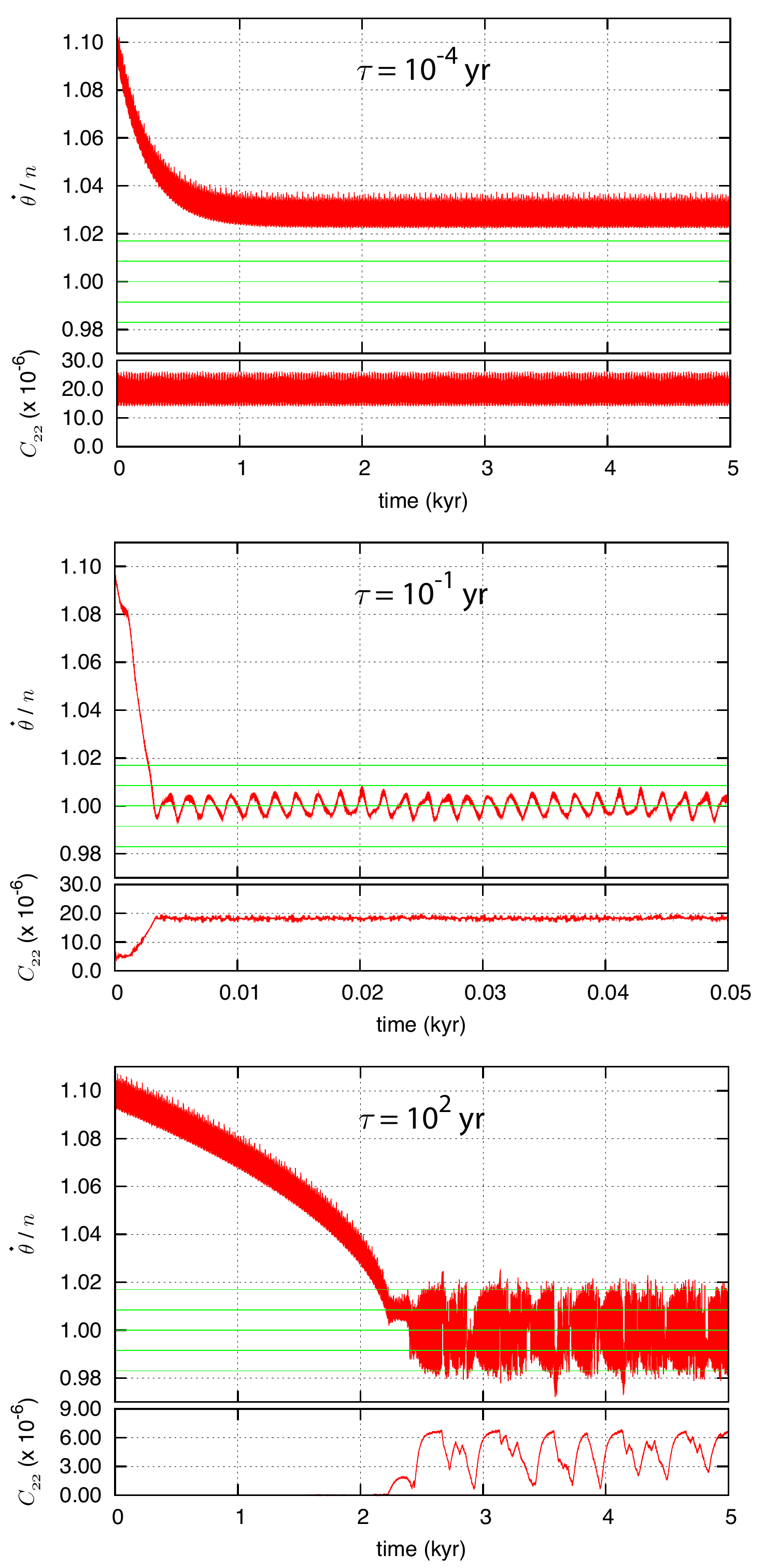}
    \caption{Examples of the spin evolution of \object{Kepler-88}\,$b$ and corresponding global $C_{22}$ value for different $\tau$ values. We use the initial conditions from Table~\ref{tab:III}, and the initial rotation period is set at 10~d. The green lines give the position of the spin-orbit resonances $n$, $n\pm\nu/2$, and $n\pm\nu$.
    }
    \label{fig:V}
  \end{figure}
}

\newcommand\tabI{
  \begin{table}
    \begin{center}
      \caption{Qualitative evolution of the spin as a function of the amplitude
      $\alpha$ and frequency $\nu$ of the perturbation.}
      \begin{tabular}{c|c|c|c}
        \hline
        & $\sigma/\nu \ll 1$ & $\sigma/\nu \sim 1$ & $\sigma/\nu \gg 1$\\
        \hline
        $\alpha \ll 1$ & \multicolumn{3}{c}{synchronous resonance only} \\
        \hline
        $\alpha \sim 1$ & 3 separated res. & chaos & modulated pendulum\\
        \hline
      \end{tabular}
      \tablefoot{The value of $\sigma$ (width of the synchronous resonance)
        depends on the $C_{22}$ coefficient of the planet (see Eq.~(\ref{eq:sigma})),
        which can be evaluated using Appendices~\ref{sec:C22} and \ref{sec:forced}.
      }
      \label{tab:I}
    \end{center}
  \end{table}
}
\newcommand\tabII{
  \begin{table}
    \begin{center}
      \caption{Parameters for \object{KOI-227}\,$b$, $c$ used in this study.}
      \begin{tabular}{cc|cc}
        \hline
        Parameter & [unit] & $b$ & $c$ \\
        \hline
        $m$ & [$M_\oplus$] &
        $11.09$ & $43.89$ \\
        $R$ & [$R_\oplus$] &
        $2.23$ & -- \\
        $\kf$ &  &
        $0.95$ & -- \\
        $\zetai$ &  &
        $0.333$ & -- \\
        $C_{22,r}$ &  &
        $1.4 \times 10^{-7}$ & -- \\
        \hline
        $a$ & [AU] &
        $0.104483724$ & $0.167413850$ \\
        $e$ & &
        $0.0756724$ & $0.0173317$ \\
        $\lambda$ & [deg] &
        $0$ & $250.4696$ \\
        $\varpi$ & [deg] &
        $-179.2092$ & $119.1518$ \\
        \hline
      \end{tabular}
      \tablefoot{The stellar mass is $0.49 M_\odot$.
        The orbital parameters are taken from \citet{nesvorny_photo_2014}.
        The reference epoch is 2,454,952.024800880921 BJD.
        For the sake of simplicity of the model, we assume the system to be coplanar.
        This solution has a $\chi^2$ of $53.3$.}
      \label{tab:II}
    \end{center}
  \end{table}
}

\newcommand\tabIII{
  \begin{table}
    \begin{center}
      \caption{Parameters for \object{Kepler-88}\,$b$, $c$ used in this study.}
      \begin{tabular}{cc|cc}
        \hline
        Parameter & [unit] & $b$ & $c$ \\
        \hline
        $m$ & [$M_\oplus$] &
        $8.7$ & $198.8$ \\
        $R$ & [$R_\oplus$] &
        $3.78$ & -- \\
        $\kf$ &  &
        $0.45$ & -- \\
        $\zetai$ &  &
        $0.25$ & -- \\
        $C_{22,r}$ &  &
        0 & -- \\
        \hline
        $a$ & [AU] &
        $0.095093133$ & $0.152955525$ \\
        $e$ & &
        $0.05593$ & $0.05628$ \\
        $i$ & [deg] &
        $0.945$ & $3.8$ \\
        $\lambda$ & [deg] &
        $6.405$ & $252.9$ \\
        $\varpi$ & [deg] &
        $90.59$ & $270.76$ \\
        $\Omega$ & [deg] &
        $270$ & $264.1$ \\
        \hline
      \end{tabular}
      \tablefoot{The stellar mass is $0.956 M_\odot$.
        The reference epoch is 2,454,954.62702 BJD.
        The orbital parameters are taken from \citet{nesvorny_king_2013}.
        In our simulations we neglect the small mutual inclination and assume the planets to be coplanar.}
      \label{tab:III}
    \end{center}
  \end{table}
}

\newcommand\tabIV{
\begin{table}
\begin{center}
\caption{Capture probabilities in spin-orbit resonances (in percent) for \object{KOI-227}\,$b$,
using different $\tau$ values. \label{tab:IV}}
\begin{tabular}{c | c | c | c | c}
\hline
\multirow{2}{*}{$\eta-n$} & \multicolumn{4}{c}{$\tau$ (yr)}  \\ &  $10^1$  & $10^2$  &  $10^3$ & $10^4$ \\
\hline
$\nu   $ &   $-$ &   $-$ &    3.8   &   2.3  \\
$\nu/2$ &   $-$ &    57.3  &   74.2  & 31.4 \\
$0$       &  100.0 &     42.7  &   22.0  & 56.7  \\
$-\nu/2$ & $-$ &    $-$ &   $-$  &  9.6  \\
\hline
\end{tabular}
\end{center}
\end{table}
}

\begin{document}

\title{Spin dynamics of close-in planets exhibiting large TTVs}
\titlerunning{Spin of TTV planets}
\author{J.-B. Delisle\inst{1,2}
  \and A. C. M. Correia\inst{2,3}
  \and A. Leleu\inst{2,4,}\thanks{CHEOPS fellow}
  \and P. Robutel\inst{2}}
\institute{Observatoire de l'Université de Genève, 51 chemin des Maillettes, 1290, Sauverny, Switzerland\\
  \email{jean-baptiste.delisle@unige.ch}
    \and ASD, IMCCE, Observatoire de Paris - PSL Research University, UPMC Univ. Paris 6, CNRS,\\
    77 Avenue Denfert-Rochereau, 75014 Paris, France
  \and CIDMA, Departamento de F\'isica, Universidade de Aveiro, Campus de Santiago,
  3810-193 Aveiro, Portugal
  \and Physikalisches Institut \& Center for Space and Habitability, Universitaet Bern, 3012 Bern, Switzerland
}

\date{\today}

\abstract{
We study the spin evolution of close-in planets in compact multi-planetary systems.
The rotation period of these planets is often assumed to be synchronous with the orbital period due to tidal dissipation.
Here we show that planet-planet perturbations can drive the spin of these planets into non-synchronous or even chaotic states.
In particular, we show that the transit timing variation (TTV) is
a very good probe to study the spin dynamics,
since both are dominated by the perturbations of the mean longitude of the planet.
We apply our model to \object{KOI-227}\,$b$ and \object{Kepler-88}\,$b$, which are both observed undergoing strong TTVs.
We also perform numerical simulations of the spin evolution of these two planets.
We show that for \object{KOI-227}\,$b$ non-synchronous rotation is possible, while for \object{Kepler-88}\,$b$ the rotation can be chaotic.}

\keywords{celestial mechanics -- planets and satellites: general}

\maketitle


\section{Introduction}
\label{sec:introduction}

The rotation of close-in planets is usually modified by tidal interactions with the central star and reaches a stationary value on time-scales
typically much shorter than the tidal evolution of orbits
\citep[e.g.][]{Hut_1981, Correia_2009}.
As long as the orbit has some eccentricity, the rotation can stay in non-synchronous configurations,
but tidal dissipation also circularizes the orbit, which ultimately results in synchronous motion (the orbital and rotation periods become equal).

Recently, \citet{leconte_asynchronous_2015} used simulations
including global climate model (GCM)
of the atmosphere of Earth-mass planets in the habitable zone of
M$-$type stars to show that these planets might be in a state
of asynchronous rotation \citep[see also][]{correia_equilibrium_2008}.
This asynchronous rotation is due to thermal tides in the atmosphere.
This same effect was also invoked to explain
the retrograde spin of Venus \citep[see][]{correia_four_2001, Correia_Laskar_2003I}.
However, for close-in planets, the gravitational tides dominate the thermal tides,
so synchronous rotation is believed to be the most likely scenario
\citep{correia_equilibrium_2008,cunha_spin_2015}.

In this paper we investigate another effect that can drive the spin of close-in planets
to asynchronous rotation, namely planetary perturbations.
\citet{correia_spin_2013} showed that in the case of co-orbital planets,
planet-planet interactions induce orbital perturbations that
can lead to asynchronous spin equilibria,
and even chaotic evolution of the spin of the planets.
The planets librate around the Lagrangian equilibrium
and have oscillations of their mean longitude that prevent the spin synchronization.

We generalize this study to other mean-motion resonances (2:1, 3:2, etc.)
for which a similar libration of the mean longitude can be observed.
We study both the resonant and the near-resonance cases.
While no co-orbital planet has yet been observed, many planets have been found
around other mean-motion resonances.
Moreover, in some cases,
strong planet-planet interactions have been observed,
for instance \object{GJ~876} using the radial velocity (RV) technique
\citep{correia_harps_2010},
or \object{Kepler-88} (also known as the King of TTVs)
using TTV technique \citep{nesvorny_king_2013}.
The TTV technique is particularly promising for such a study
since the amplitude and period of the observed TTVs
provide a direct measure of the amplitude and period of the orbital libration
which affects the spin evolution.

In Sect.~\ref{sec:model}, we derive the equations of motion for the spin
of a planet that is perturbed by another planet.
In Sect.~\ref{sec:ttvs} we show
that both the spin dynamics and the TTVs
are dominated by the perturbations of the mean longitude of the planet,
and the coefficients of their Fourier series are related to each other.
In Sect.~\ref{sec:obs}, we apply our analytical modelling
to some observed planets showing significant TTVs
and perform numerical simulations.
Finally, we discuss our results in Sect.~\ref{sec:discussion}.

\section{Spin dynamics}
\label{sec:model}

We consider a system consisting of a central star with mass $m_0$, and two companion planets with masses $m_1$ and $m_2$, such that $m_1, m_2 \ll m_0$.
We study the spin evolution of one of the planets, either the inner one or the outer one.
The subscript 1 always refers to the inner planet, 2 refers to the outer one,
and no subscript refers to the planet whose spin evolution is studied.
For simplicity, we assume coplanar orbits and low eccentricities for both planets.
We assume that the spin axis of the considered planet
is orthogonal to the orbital plane (which corresponds to zero obliquity)
\footnote{Tidal effects drive the obliquity of the planets to zero degrees
\citep[e.g.][]{correia_longterm_2003,boue_complete_2016},
so we expect that Kepler planets whose spin has been driven close to the synchronous rotation
also have nearly zero obliquity.}.

We introduce $\theta$, the rotation angle of the planet with respect to an inertial line,
whose evolution is described by \citep[e.g.][]{murray_solar_1999}
\begin{equation}
  \label{eq:ddtheta}
  \ddot{\theta} = - \frac{6 \C22i}{\zetai} \frac{\mu}{r^3} \sin 2(\theta - f) \ ,
\end{equation}
where
$\mu = \G (m_0+m)$ ($\G$ being the gravitational constant),
$r$ is the distance between the planet and the star (heliocentric coordinates),
$f$ is the true longitude of the planet,
$\C22i$ is the Stokes gravity field coefficient that measures
the asymmetry in the equatorial axes of the planet,
and $\zetai$ is the inner structure coefficient that measures
the distribution of mass in the planet's interior (see Appendix~\ref{sec:C22}).

Expanding expression (\ref{eq:ddtheta}) in Fourier series of the mean longitude, we obtain
\begin{equation}
  \label{eq:ddtheta2}
  \ddot{\theta} = - \frac{6 \C22i}{\zetai} \frac{\mu}{a^3}
  \sum_{k\in\mathbb{Z}} X_k^{-3,2}(e) \sin (2\theta - k\lambda + (k-2)\varpi) \ ,
\end{equation}
where $a$, $e$, $\lambda$, $\varpi$ are
the semi-major axis, the eccentricity, the mean longitude,
and the longitude of periastron of the planet, respectively.
We note that the Hansen coefficient $X_k^{-3,2} (e)$ is of order $|k-2|$ in eccentricity.
Thus, the leading term in this expansion corresponds to $k=2$,
and we have $X_2^{-3,2}(e) = 1 + O(e^2)$.
If we only keep this leading term, Eq.~(\ref{eq:ddtheta2}) simplifies as
\begin{equation}
  \label{eq:ddtheta3}
  \ddot{\theta} \approx - \frac{6 \C22i}{\zetai} \frac{\mu}{a^3} \sin 2(\theta - \lambda) \ .
\end{equation}

If the planet remains unperturbed and follows a Keplerian orbit,
the semi-major axis is constant and the mean longitude is given by
\begin{equation}
  \label{eq:longkeplerian}
  \lambda = \lambda_0 + n t,
\end{equation}
where $n$ is the constant mean motion of the planet.
Introducing
\begin{eqnarray}
  \label{eq:defgamma}
  \gamma &=& \theta - \left(\lambda_{0} + n t\right),\\
  \label{eq:sigma}
  \sigma &=& n \sqrt{12 \, \C22i / \zetai} \ ,
\end{eqnarray}
we can rewrite Eq.~(\ref{eq:ddtheta3}) as
\begin{equation}
  \label{eq:ddgammaKep}
  \ddot{\gamma} \approx - \frac{\sigma^2}{2} \sin 2\gamma \ ,
\end{equation}
which is the equation of a simple pendulum, with a stable equilibrium point at $\gamma = 0$.
This equilibrium corresponds to the exact synchronization ($\dot{\theta} = n$).
The parameter $\sigma$ (Eq.\,(\ref{eq:sigma})) measures the width of the synchronous resonance, and also corresponds to the libration frequency at exact resonance.

We now consider planet-planet interactions that disturb the orbits, in particular $a_i$ and $\lambda_i$.
The action canonically conjugated with $\lambda_i$ is the circular angular momentum of planet $i$,
$\Lambda_i = \beta_i\sqrt{\mu_i a_i}$, where $\beta_i = m_i m_0/(m_0+m_i)$.
We also introduce the angular momentum deficit (AMD) of the planets,
$D_i=\Lambda_i\left(1-\sqrt{1-e_i^2}\right)$
which are the actions canonically conjugated with $\varpi_i$.
The Hamiltonian of the three-body problem reads
\begin{equation}
  \label{eq:hamil}
  \H = \H_0(\Lambda_i) + \H_1(\Lambda_i,\lambda_i,D_i,\varpi_i),
\end{equation}
where $\H_0$ and $\H_1$ are the Keplerian and perturbative part of the Hamiltonian.
We have
\begin{equation}
  \H_0 = -\frac{\mathcal{G} m_0 m_1}{2 a_1} -\frac{\mathcal{G} m_0 m_2}{2 a_2}
  = -\frac{\mu_1^2\beta_1^3}{2\Lambda_1^2} - \frac{\mu_2^2\beta_2^3}{2\Lambda_2^2},
\end{equation}
and $\H_1$ is of order one in planet masses (over star mass).
The equations of motion read
\begin{eqnarray}
  \dot{\lambda}_i &=& \frac{\partial \H}{\partial \Lambda_i}
  = \frac{\mu_i^2\beta_i^3}{\Lambda_i^3} + \frac{\partial \H_1}{\partial \Lambda_i},\\
  \dot{\Lambda}_i &=& - \frac{\partial \H}{\partial \lambda_i}
  = - \frac{\partial \H_1}{\partial \lambda_i}.
\end{eqnarray}
However, it can be shown that the perturbative part has a weak dependency on $\Lambda_i$
\citep[e.g.][]{delisle_resonance_2014}.
At leading order in eccentricity, we have
\begin{equation}
  \label{eq:dla}
  \dot{\lambda}_i \approx \frac{\mu_i^2\beta_i^3}{\Lambda_i^3}
  \approx n_i\left(\frac{\Lambda_i}{\Lambda_{i,0}}\right)^{-3}
  \approx n_i - 3 n_i \frac{\Lambda_i-\Lambda_{i,0}}{\Lambda_{i,0}}.
\end{equation}
We assume that the orbital motion is not chaotic and thus quasi-periodic.
We introduce the quasi-periodic decomposition of $\Lambda$
\begin{equation}
  \label{eq:Lat}
  \Lambda = \Lambda_{0} \left( 1 + \sum_j A_j \cos(\nu_j t + \phi_j) \right),
\end{equation}
where $\nu_j$ can be any combination of the frequencies of the system.
For a resonant system (see Appendix~\ref{sec:resonant-case}),
the frequency of libration in
the resonance is the dominant term of the decomposition.
For a system that is close to resonance ($k_2$:$k_1$)
but outside of it (see Appendix~\ref{sec:near-resonance-case}),
the dominant frequency is the frequency of circulation $k_2 n_2 - k_1 n_1$.
However, for this computation we can keep things general and use the
generic decomposition of Eq.~(\ref{eq:Lat}).
We replace it in Eq.~(\ref{eq:dla}) to obtain the evolution of the mean longitude
\begin{equation}
  \dot{\lambda} \approx n - 3 n \sum_j A_j \cos(\nu_j t + \phi_j).
\end{equation}
We thus have
\begin{eqnarray}
  \lambda &\approx& \lambda_{0} + n t - \sum_j 3A_j\frac{n}{\nu_j}
  \left(\sin(\nu_j t + \phi_j)-\sin\phi_j\right)\nonumber\\
  \label{eq:lat}
  &=& \lambda_{0} + n t - \sum_j \alpha_j
  \left(\sin(\nu_j t + \phi_j)-\sin\phi_j\right),
  \label{eq:lambda}
\end{eqnarray}
where the coefficients
\begin{equation}
  \label{eq:defalpha}
  \alpha_j = 3 A_j n/\nu_j
\end{equation}
have the dimension of angles
and correspond to the amplitudes of each term in the decomposition.
For the sake of simplicity of the computations,
we assume that these amplitudes remain small,
but in principle they could reach $180^\circ$.
With this approximation, we have at first order in $\alpha_j$
\begin{equation}
  \label{eq:sintheta}
  \sin 2(\theta-\lambda) \approx \sin 2\gamma + 2 \sum_j \alpha_j \sin(\nu_j t + \phi_j) \cos 2\gamma,
\end{equation}
with
\begin{equation}
  \label{eq:defgammab}
  \gamma = \theta - \left(\lambda_{0} + n t + \sum_j \alpha_j \sin\phi_j\right).
\end{equation}
From Eq.~(\ref{eq:Lat}) we deduce
\begin{eqnarray}
  \label{eq:smavar}
  \frac{1}{a^3} &\approx& \frac{1}{a_{0}^3} \left( 1 - 6 \sum_j A_j \cos(\nu_j t + \phi_j) \right)\nonumber\\
  \label{eq:act}
  &\approx& \frac{1}{a_{0}^3} \left( 1 - 2 \sum_j \alpha_j \frac{\nu_j}{n} \cos(\nu_j t + \phi_j) \right)
\end{eqnarray}
and finally, using the expressions of Eqs.~(\ref{eq:sintheta}),~(\ref{eq:act}) in
Eq.~(\ref{eq:ddtheta3}),
we obtain
\begin{eqnarray}
  \ddot{\gamma} \approx
    -\frac{\sigma^2}{2}
    \Bigg[\!\!\!\!&&\!\!\!\!\sin 2\gamma\nonumber\\
    \label{eq:ddgamma}
    &&\!\!\!\!+\sum_j \alpha_j \left(1-\frac{\nu_j}{n}\right) \sin 2\left(\gamma + \frac{\nu_j t + \phi_j}{2}\right)\\
    &&\!\!\!\!+ \sum_j \alpha_j \left(1+\frac{\nu_j}{n}\right) \sin 2\left(\gamma - \frac{\nu_j t + \phi_j + \pi}{2}\right)\Bigg].\nonumber
\end{eqnarray}
The term ($\sin 2\gamma$) corresponding to the synchronous resonance still exists.
However, for each frequency appearing in the quasi-periodic decomposition of the
perturbed orbital elements (see Eqs.~(\ref{eq:Lat}),~(\ref{eq:lat})),
two terms appear (at first order in $\alpha_j$) in Eq.~(\ref{eq:ddgamma}),
corresponding to a sub-synchronous resonance ($\dot{\theta}=n-\nu_j/2$),
and a super-synchronous resonance ($\dot{\theta}=n+\nu_j/2$).
This splitting of the synchronous resonance was found in the case of co-orbital planets
in \citet{correia_spin_2013}.
We note that if we do not neglect the non-leading terms in the Fourier expansion (Eq.~(\ref{eq:ddtheta2}),
the classical spin-orbit resonances ($\dot\theta = kn/2$) appear,
as do new resonances of the form $\dot\theta = kn/2 \pm \nu_j/2$.
Moreover, if the amplitudes $\alpha_j$ are not small,
the series should be developed at a higher degree in $\alpha_j$,
and resonances of the type $\dot\theta = kn/2 \pm l \nu_j/2$
would appear \citep[see][for the co-orbital case]{leleu_rotation_2015}.

Different dynamical regimes can be observed depending on the values of
$\alpha$, $\nu$, and $\sigma$.
In particular, a chaotic evolution of the spin is expected when the separation
between two resonances is of the order of the width of these resonances
\citep{chirikov_universal_1979}.
Table~\ref{tab:I} describes qualitatively the dynamics of the spin of the planet,
as a function of the amplitude ($\alpha$) and frequency ($\nu$) of the perturbing term.

\tabI

\section{TTV as a probe for the spin dynamics}
\label{sec:ttvs}

The TTV of a planet is a very good probe that can be used to estimate
the main frequencies appearing in the quasi-periodic decomposition of Eqs.~(\ref{eq:Lat}),~(\ref{eq:lat}), and the associated amplitudes.
As we did for the computation of the spin evolution,
we consider coplanar planets with low eccentricities.
We take the origin of the longitudes as the observer direction such that the
transit occurs when $f = 0$ ($f$ being the true longitude of the planet).
The true longitude can be expressed as a Fourier series of the mean longitude
\begin{equation}
  \expo{\im f} = \sum_{k\in\mathbb{Z}} X_k^{0,1} (e) \expo{\im (k \lambda+(1-k)\varpi)}
\end{equation}
where the Hansen coefficient $X_k^{0,1}$ is of order $|k-1|$ in eccentricity.
For the sake of simplicity, we only keep the leading order term $f \approx \lambda$.
We thus have (see Eq.~(\ref{eq:lat}))
\begin{equation}
  f \approx \lambda_{0} + n t - \sum_j \alpha_j
  \left(\sin(\nu_j t + \phi_j)-\sin\phi_j\right)
\end{equation}
We introduce $t_k$, the time of the $k$-th transit.
We have $f(t_k) = k 2\pi$, thus the transit timing variations are given by
\begin{equation}
  \label{eq:ttvs}
  \mathrm{TTV}_k = t_k - k P \approx
  \sum_j \frac{\alpha_j}{n}\left(\sin(\nu_j t_k + \phi_j)-\sin\phi_j\right)
  - \frac{\lambda_{0}}{n},
\end{equation}
where $P = 2\pi/n$ is the orbital period.
Therefore, the quasi-periodic decomposition of the TTV signal directly
provides the amplitudes $\alpha_j$ and frequencies $\nu_j$ that we need
in order to analyze the spin evolution.
We are mainly interested in resonant and near-resonant systems.
In these cases, one term is leading the expansion with a period much longer
than the orbital period (libration period for the resonant case, and circulation period for the near-resonant case,
see Appendices~\ref{sec:near-resonance-case},~\ref{sec:resonant-case}).
For these long periods, the amplitudes and frequency are well determined as long as
the number of transits is sufficient to cover the whole period.
In particular, there are no sampling/aliasing issues that could arise for periods that are close to the orbital period.

For the sake of simplicity we assume that one term is leading the TTVs,
such that (see Eq.~(\ref{eq:ttvs}))
\begin{equation}
  \label{eq:ttvoneterm}
  \mathrm{TTV}_k \approx \frac{\alpha}{n} \left(\sin(\nu t_k + \phi)-\sin(\phi)\right) - \frac{\lambda_{0}}{n},
\end{equation}
and (see Eq.~(\ref{eq:ddgamma}))
\begin{eqnarray}
  \label{eq:ddgammaoneterm}
  \ddot{\gamma}
  \approx
  -\frac{\sigma^2}{2}
  \bigg[ \sin 2\gamma \!\!\!\!&+&\!\!\!\! \alpha \sin 2\left(\gamma + \frac{\nu t + \phi}{2}\right)
  \nonumber \\
  &+&\!\!\!\!  \alpha \sin 2\left(\gamma - \frac{\nu t + \phi +\pi}{2}\right) \bigg] \ ,
\end{eqnarray}
where we assume ($\nu \ll n$).
The most interesting systems
for our study (see Table~\ref{tab:I})
are those for which $\alpha$ is not negligible,
such that the width of the resonances at $\dot{\theta}=n\pm\nu/2$ is not negligible
compared to the synchronous resonance.
Such systems can be locked in sub/super-synchronization or
even show chaotic evolution of the spin \citep[see][for the co-orbital case]{correia_spin_2013}.
The TTVs provide a determination of the parameters $\alpha$ and $\nu$ (and of the phase $\phi$).

\section{Application to planets with TTVs}
\label{sec:obs}

In this section we apply the results obtained in Sects.~\ref{sec:model} and~\ref{sec:ttvs} to real planetary systems that show large TTVs, and we perform numerical simulations
in the conservative case (Sect.~\ref{sec:conssim}) and dissipative case (Sect.~\ref{sec:numsims}).
We have chosen two examples: \object{KOI-227}\,$b$, which is trapped in a mean-motion resonance, and \object{Kepler-88}\,$b$, which is near (but not trapped in) a mean-motion resonance.
In both case the perturber is not observed to transit but is inferred from the TTV signal.
In addition, \object{KOI-227}\,$b$ is considered a rocky planet with a permanent equatorial asymmetry $C_{22} \ne 0 $ (Eq.\,(\ref{eq:sigma})), while \object{Kepler-88}\,$b$ is a gaseous planet
for which the $C_{22}$ value is likely very close to zero \citep{Campbell_Synnott_1985}.

\subsection{Conservative evolution}
\label{sec:conssim}
\subsubsection{KOI-227\,$b$ (rocky planet, in resonance)}

\tabII

\figI

\object{KOI-227} hosts at least two planets
\citep[see][]{nesvorny_photo_2014},
but only one (\object{KOI-227}\,$b$) is known to transit.
This planet has a radius of $2.23\ R_\oplus$,
a period of about 18~d,
and TTVs with an amplitude of at least 10~hr
have been observed
\citep[see][]{nesvorny_photo_2014}.
In terms of angular amplitude (see Eq.~(\ref{eq:lat})),
this corresponds to $\alpha \gtrsim 8^\circ$.
The main TTV period is about 4.5~yr,
thus $\nu/n \approx 0.011$.

Since the pertubing planet is not detected directly,
the orbital parameters of the system
cannot be completely solved for, due to degeneracies \citep[see][]{nesvorny_photo_2014}.
Three possible families of solutions have been proposed by \citet{nesvorny_photo_2014},
corresponding to an outer 2:1 or 3:2 resonance or an inner 3:2 resonance
between the observed planet and the perturber.
For these three solutions, the planets must stay inside the resonance.
The outer 2:1 configuration is favoured by the data but the two other configurations
cannot be ruled out \citep[see][]{nesvorny_photo_2014}.
The best-fitting solution is the 2:1 configuration
($\chi^2 = 37.6$).
The mass of \object{KOI-227}\,$b$ is $37.5\ M_\oplus$ for this solution,
and its density would thus be $18.6\ \mathrm{g\ cm^{-3}}$,
which seems very high.
However, the orbital parameters, and masses are not very well constrained.
As an example, we refitted the orbital parameters of the system,
using the same TTV data as \citet{nesvorny_photo_2014},
but imposing the density of \object{KOI-227}\,$b$ to be the same as that of the Earth.
The mass of \object{KOI-227}\,$b$ is thus set to $11.09\ M_\oplus$.
The obtained solution has a $\chi^2$ of 39.1,
which is still better than the best-fitting
solution in other resonances
\citep[$\chi^2=51.5$ for the outer 3:2 resonance,
and $\chi^2=82.2$ for the inner 3:2 resonance, see][]{nesvorny_photo_2014}.

For this illustration, we adopted the mass of $11.09\ M_\oplus$ for \object{KOI-227}\,$b$.
We also imposed the system to be coplanar in order to simplify the problem.
We thus refitted the model imposing zero inclination between the planets,
which also provides a good fit to the data ($\chi^2=53.3$).
The obtained solution is given in Table~\ref{tab:II}.

Using the mass ($11.09\ M_\oplus$) and radius
($2.23\ R_\oplus$) of the planet,
we estimate its permanent deformation (see Appendix~\ref{sec:C22})
\begin{equation}
  C_{22,r} \approx 1.4 \times 10^{-7}.
  \label{170207a}
\end{equation}
This corresponds to $\sigma/n \approx 2.2 \times 10^{-3}$ and $\sigma/\nu \approx 0.20$,
which means that the sub/super-synchronous
resonances ($\dot\theta = n \pm \nu/2$)
are well separated from the synchronous resonance,
and that the planet could be locked in any of these resonances.

In addition to the permanent deformation, the tidal deformation could also play
an important role in the spin dynamics of the planet.
In particular, if the planet is captured in one of the spin-orbit resonances,
the $C_{22}$ increases due to the tidal deformation.
We estimate the maximum deformation of the planet
(see Appendices~\ref{sec:C22} and \ref{sec:forced})
in the synchronous resonance
\begin{equation}
  C_{22, \text{sync.}} \approx 1.9 \times 10^{-6} \ ,
    \label{170208a}
\end{equation}
and in the sub/super-synchronous resonances
\begin{equation}
  C_{22, \text{sub/super}} \approx 6.6 \times 10^{-7} \ .
  \label{170208b}
\end{equation}
These values correspond to
$\sigma/n \approx 8.3\times 10^{-3}$ and $\sigma/\nu \approx 0.77$ (synchronous resonance),
and
$\sigma/n \approx 4.9\times 10^{-3}$ and $\sigma/\nu \approx 0.45$ (sub/super-synchronous resonances).
As $\sigma/\nu$ approaches unity,
the resonances get closer and closer,
which may induce a chaotic evolution of the spin.

To study in more details the spin dynamics in these different resonances,
we perform numerical simulations of the spin in the conservative case.
We substitute in Eq.~(\ref{eq:ddtheta})
the orbital solution given by a classical N-body integrator,
and integrate it to obtain the evolution of the rotation angle ($\theta$).
Figure~\ref{fig:I} shows a frequency analysis
\citep[using the NAFF algorithm, see][]{laskar_secular_1988,laskar_chaotic_1990,laskar_frequency_1993}
of the spin of \object{KOI-227}\,$b$
in the conservative case,
and assuming $C_{22}=1.4 \times 10^{-7}$ (permanent deformation),
$6.6\times 10^{-7}$ (maximum deformation in the sub/super-synchronous resonances),
and $1.9\times 10^{-6}$ (maximum deformation in the synchronous resonance).

For $C_{22}=1.4 \times 10^{-7}$ (see Fig.~\ref{fig:I} top),
we observe that the synchronous resonance is stable,
as are the main super/sub-synchronous resonances ($\eta = n \pm \nu/2$).
The widths of these three resonances are comparable,
which indicates that the capture probability in any of these resonances
should be similar.
For $C_{22}=6.6 \times 10^{-7}$ (see Fig.~\ref{fig:I} middle),
we observe that the three resonances are surrounded by a large chaotic area.
However, a stable region is still visible in each of the three resonances.
This means that the sub/super-synchronous resonances remain stable
even if the tidal deformation increases to its maximum value after the resonant capture.
Finally, for $C_{22}=1.9 \times 10^{-6}$ (see Fig.~\ref{fig:I} bottom),
we observe a very large chaotic area
that encompasses all the resonances.
We still observe three areas of stability, which correspond
to the synchronous resonance and to the
second-order sub/super-synchronous resonances $\eta = n \pm \nu$).
Stable capture in the synchronous resonance is thus still possible.

We conclude that in the case of \object{KOI-227}\,$b$,
stable captures in the synchronous,
and sub/super-synchronous resonances ($\eta = n \pm \nu/2$)
are all possible, and should have comparable probabilities.
However, for the highest $C_{22}$ value some chaotic evolution can be expected before the rotation enters a stable island.

\subsubsection{Kepler-88\,$b$ (gaseous planet, close to resonance)}
\label{KOI142nrc}

\tabIII

\figII

\object{Kepler-88}\,$b$, also referred to as \object{KOI-142}\,$b$, or as the King of TTVs,
is the planet exhibiting the largest TTV known today.
The TTV amplitude is $\alpha\approx16^\circ$
\citep[amplitude of 12~hr, compared with an orbital period of 10.95~d, see][]{nesvorny_king_2013},
and the TTV period is about 630~d,
thus we have
$\nu/n \approx 0.017$.
As is true for \object{KOI-227}\,$b$,
the perturber is not directly observed.
However, the TTV signal and the transit duration variation (TDV)
are sufficient in this case to obtain a unique orbital solution \citep[see][]{nesvorny_king_2013}.
We reproduce in Table~\ref{tab:III}
the orbital parameters obtained by \citet{nesvorny_king_2013}.
We note that the mutual inclination is very small (about $3^\circ$),
and we neglect it in the following.

The bulk density of \object{Kepler-88}\,$b$ is about $0.87\ \mathrm{g.cm}^{-3}$
\citep{nesvorny_king_2013}, which means that this planet is mainly gaseous.
Its permanent deformation is thus probably very small and we neglect it
($C_{22,r}\approx 0$).
However, the tidally induced deformation of the planet, if it is captured in
a resonance, is not negligible (see Appendices~\ref{sec:C22} and \ref{sec:forced})
\begin{eqnarray}
  C_{22,\text{sync.}} &\approx& 6.7 \times 10^{-6} \ ,    \label{170217a} \\
  C_{22,\text{sub/super}} &\approx& 2.2\times 10^{-6} \ .
\end{eqnarray}
This corresponds to
$\sigma/n \approx 0.018$ and $\sigma/\nu \approx 1.0$
(maximum deformation in the synchronous resonance),
and $\sigma/n \approx 0.010$ and $\sigma/\nu \approx  0.59$
(maximum deformation in the sub/super-synchronous resonances).

Figure~\ref{fig:II} shows the spin dynamics of \object{Kepler-88}\,$b$ in
the conservative case and with both estimates of the deformation.
For $C_{22}=2.2 \times 10^{-6}$ (see Fig.~\ref{fig:II} top),
we observe that a large chaotic area is surrounding the synchronous
and super/sub-synchronous resonances ($\eta = n \pm \nu/2$).
A stable area is visible at the centre of the synchronous resonance,
but not in the super/sub-synchronous resonances.
Therefore, permanent capture in non-synchronous resonances is not possible.
For $C_{22}=6.7 \times 10^{-6}$ (see Fig.~\ref{fig:II}, bottom),
a very small island of stability is still present in
the synchronous resonance.
Therefore, stable capture in the synchronous resonance should be possible but might be
difficult to achieve.

\subsection{Numerical simulations with tidal dissipation}
\label{sec:numsims}

Tidal interactions with the star have a double effect on the planet:
deformation and dissipation.
The deformation occurs because the mass distribution inside the planet adjusts to the tidal potential.
The dissipation occurs because this adjustment is not instantaneous, so there is a lag between the perturbation and the maximum deformation.
As seen in Sect.~\ref{sec:conssim}, the deformation is very important,
since different values of the $C_{22}$ can
have very different implications for the spin dynamics.

In this section we also take into account the dissipative part of the tidal effect,
which slowly modifies the spin rotation rate of the planet,
and might drive it into the different configurations described in Sect.~\ref{sec:conssim}.
In order to get a comprehensive picture of the spin dynamics of the considered planets,
and especially to estimate capture probabilities in the different spin-orbit resonances,
we run numerical simulations that take into account both
the tidal deformation and the tidal dissipation.

Viscoelastic rheologies have been shown to reproduce the main features of tidal effects \citep[for a review of the main models, see][]{Henning_etal_2009}.
One of the simplest models of this kind is to consider that the planet behaves like a Maxwell material, which is represented by a purely viscous damper and a purely elastic spring connected in series \cite[e.g.][]{Turcotte_Schubert_2002}.
In this case, the planet can respond as an elastic solid or as a viscous fluid, depending on the frequency of the perturbation.
The response of the planet to the tidal excitation is modelled by the parameter $\tau$, which corresponds to the relaxation time of the planet\footnote{$\tau = \tau_v + \tau_e$, where $\tau_v$ and $\tau_e$ are the viscous (or fluid) and Maxwell (or elastic) relaxation times, respectively. For simplicity, in this paper we consider $\tau_e=0$, since this term does not contribute to the tidal dissipation \citep[for more details, see][]{Correia_etal_2014}.}.

We adopt here a Maxwell viscoelastic rheology using a differential equation for the gravity field coefficients \citep{Correia_etal_2014}.
This method tracks the instantaneous deformation of the planet, and therefore allows us to correctly take into account the gravitational perturbations from the companion body.
The complete equations of motion governing the orbital evolution of the system in an astrocentric frame are \citep{Rodriguez_etal_2016}
\begin{eqnarray}\label{mov1}
\ddot{\vec{r}}_1&=&-\frac{\mu_1}{r_1^3}\vec{r}_1+ \mathcal{G} m_2\left(\frac{\vec{r}_2-\vec{r}_1}{|\vec{r}_2-\vec{r}_1|^3}-\frac{\vec{r}_2}{r_2^3}\right) + \vec{f}   \ ,
\end{eqnarray}
\begin{eqnarray}\label{mov2}
\ddot{\vec{r}}_2&=&-\frac{\mu_2}{r_2^3}\vec{r}_2+ \mathcal{G} m_1\left(\frac{\vec{r}_1-\vec{r}_2}{|\vec{r}_1-\vec{r}_2|^3}-\frac{\vec{r}_1}{r_1^3}\right) + \frac{\mathcal{G} m_1}{\mu_1} \vec{f} \ ,
\end{eqnarray}
where $\vec{r}_i$ is the position vector of the planet $i$ (astrocentric coordinates),
$\vec{f}$ is the acceleration arising from the potential created by the deformation of the inner planet  \citep{Correia_etal_2014}
\begin{eqnarray}\label{fdef}
\vec{f}&=&-\frac{3\mu_1 R^2}{2r_1^5}J_2\vec{r}_1 \\
&&-\frac{9\mu_1 R^2}{r_1^5}\left[C_{22}\cos2(\theta-f_1)-S_{22}\sin2(\theta-f_1)\right]\vec{r}_1\nonumber\\
&&+\frac{6\mu_1 R^2}{r_1^5}\left[C_{22}\sin2(\theta-f_1)+S_{22}\cos2(\theta-f_1)\right]\vec{k}\times\vec{r}_1 \nonumber,
\end{eqnarray}
and $\vec{k}$ is the unit vector normal to the orbital plane of the inner planet.
The torque acting to modify the inner planet rotation is
\begin{equation}\label{torque}
\ddot{\theta}=-\frac{6\G m_0}{\zeta r_1^3}\left[C_{22}\sin2(\theta-f_1)+S_{22}\cos2(\theta-f_1)\right] .
\end{equation}

The inner planet is deformed under the action of self rotation and tides.
Therefore, the gravity field coefficients can change with time
as the shape of the planet is continuously adapting to the equilibrium figure.
According to the Maxwell viscoelastic rheology, the deformation law for these coefficients is given by \citep{Correia_etal_2014}
\begin{eqnarray}\label{max1}
&&J_2+\tau\dot{J}_2 = \kf \frac{\dot{\theta}^2R^3}{3\G m_1} + \kf \frac{m_0}{2m_1}\left(\frac{R}{r_1}\right)^3 \ ,\nonumber\\
&&C_{22}+\tau\dot{C}_{22}  = C_{22,r} + \frac{\kf}{4}\frac{m_0}{m_1}\left(\frac{R}{r_1}\right)^3\cos2(\theta-f_1)  \ ,\\
&&S_{22}+\tau\dot{S}_{22} = -\frac{\kf}{4}\frac{m_0}{m_1}\left(\frac{R}{r_1}\right)^3\sin2(\theta-f_1) \ , \nonumber
\end{eqnarray}
where $\kf$ is the fluid second Love number for potential.
The relaxation times $\tau$ are totally unknown for exoplanets,
but if the tidal quality dissipation $Q-$factor can be estimated,
then an equivalent $\tau$ can be obtained \citep[see][]{Correia_etal_2014}.
To cover all possible scenarios,
in our numerical simulations we adopt a wide spectrum of $\tau$ values:
$\log_{10} \tau_\mathrm{yr} = -5$ to~4, with step 1.

\subsubsection{KOI-227\,$b$ (rocky planet, in resonance)}

For rocky planets, such as the Earth and Mars, we have $Q=10$ \citep{Dickey_etal_1994} and $Q=80$ \citep{Lainey_etal_2007}, respectively.
We then compute for the Earth $\tau=1.6$~d, and for Mars $\tau=14.7$~d.
However, in the case of the Earth, the present $Q-$factor is dominated by the oceans, the Earth's solid body $Q$ is estimated to be 280 \citep{Ray_etal_2001}, which increases the relaxation time by more than one order of magnitude ($\tau=46$~d).
Although these values provide a good estimation for the average present dissipation ratios, they appear to be inconsistent with the observed deformation of the planets.
Indeed, in the case of the Earth, the surface post-glacial rebound due to the last glaciation about $10^4$~years ago is still going on, suggesting that the Earth's mantle relaxation time is something like $\tau=4400$~yr \citep{Turcotte_Schubert_2002}.
Therefore, we conclude that the deformation timescale of rocky planets can range from a few days up to thousands of years.

In our numerical simulations we use the initial conditions from Table~\ref{tab:II}.
The initial rotation period is set at 15.5~d, which corresponds to $\dot \theta / n \approx 1.14 $.
Since the libration frequency of \object{KOI-227}\,$b$ is $\nu/n \approx 0.011$, the initial rotation rate is completely outside the resonant area, which is bounded by $\dot \theta/n \le 1 + \nu/n \approx 1.01$ (Fig.\,\ref{fig:I}).
For any $\tau$ value, the rotation rate of the planet decreases due to tidal effects until it approaches this area, where multiple spin-orbit resonances are present.

Capture in resonance is a stochastic process.
Therefore, for each $\tau$ value we ran 1,000 simulations
with slightly different initial rotation rates.
The step $\Delta \dot \theta$ between each initial condition was chosen such that the moment at which each simulation crosses the resonance spreads equally over one eccentricity cycle.
In Figure~\ref{fig:III} we show some examples of evolution for different $\tau$ values.
The evolution timescale changes with $\tau$ because the $Q-$factor was also modified.

\figIII

For $\tau \le 10^{-2}$~yr, the spin is in the low frequency regime ($\tau n \ll 1$), which is usually known as the viscous or linear \citep[e.g.][]{Singer_1968, Mignard_1979}.
As a consequence, the rotation rate evolves into pseudo-synchronous equilibrium \citep[e.g.][]{Correia_etal_2014}
\begin{equation}
\dot \theta/n = 1 + 6 e^2 + {\cal O}(e^4) \ .
\label{170207b}
\end{equation}
The average eccentricity of \object{KOI-227}\,$b$'s orbit is 0.057, which gives a value
for the pseudo equilibrium of $\langle \dot \theta / n \rangle \approx 1.019$\footnote{The average rotation rate for an oscillating eccentricity is actually given by a value slightly higher than that obtained with expression (\ref{170207b}) using the average value of the eccentricity \citep[see][]{Correia_2011}.}.
This value is already very close to the synchronous resonance,
but since the libration width is $\sigma/n \approx 0.002$ (Eq.\,(\ref{170207a})),
we are still outside the resonant area.
Therefore, in this regime initial prograde rotations never cross any spin-orbit resonances.

For $10^{-1} \le \tau \le 10^1$~yr, the spin is in a transition of frequency regime ($\tau n  \sim 1$).
The rotation rate still evolves into a pseudo-synchronous equilibrium, but its value is below that provided by expression (\ref{170207b}), and lies inside the libration width of the synchronous resonance \citep[see Fig.~\ref{fig:IV} in][]{Correia_etal_2014}.
As a result, for these $\tau$ values the spin is always captured in the synchronous resonance.
Capture in higher order spin-orbit resonances is possible but with a very low probability ($< 1\%$);
we did not obtain any examples in our simulations.

For $\tau \ge 10^2$~yr, the spin is in the high frequency regime ($\tau n \gg 1$).
In this regime the tidal torque has multiple equilibria that coincide with the spin-orbit resonances \citep[see Fig.~4 in][]{Correia_etal_2014}.
Therefore, capture in the asynchronous higher order resonances becomes a real possibility.
In Table~\ref{tab:IV} we list the final distribution in the different resonances for each $\tau$ value.
We observe that as $\tau$ increases, the number of captures in higher order resonances also increases.
Indeed, for high $\tau$ values, the $C_{22}$ is able to retain its tidal deformation (Eqs.\,(\ref{170208a}) and (\ref{170208b})) for longer periods of time, increasing the libration width of the individual resonances.
When the $C_{22}$ reaches its maximum tidal deformation, some individual resonances overlap, which results in chaotic motion around these resonances, including synchronous resonance (Fig.\,\ref{fig:I}).
An interesting consequence is that for $\tau = 10^4$~yr, the sub-synchronous resonance can be reached after some wandering in this chaotic zone.
In Figure~\ref{fig:IV} we show four examples of capture in each resonance for this $\tau$ value.

\tabIV

\figIV

\subsubsection{Kepler-88\,$b$ (gaseous planet, close to resonance)}

For gaseous planets we have $Q\sim10^4$ \citep{Lainey_etal_2009, Lainey_etal_2012}, which gives $\tau$ values of a few minutes assuming that most of the dissipation arises in the convective envelope.
However, the cores of these planets also experience tidal effects, which in some cases can be equally strong \citep{Remus_etal_2012a, Guenel_etal_2014}.
In addition, other tidal mechanisms, such as the excitation of inertial waves, are expected to take place, which also enhance the tidal dissipation \citep[e.g.][]{Ogilvie_Lin_2004, Favier_etal_2014}.
Therefore, the full deformation of gaseous planets may be of the order of a few years or even decades \citep[for a review see][]{Socrates_etal_2013}.

In our numerical simulations we use the initial conditions from Table~\ref{tab:III}.
The initial rotation period is set at 10~d, which corresponds to $\dot \theta / n \approx 1.1 $.
Since the libration frequency of \object{Kepler-88}\,$b$ is $\nu/n \approx 0.017$, the initial rotation rate is completely outside the resonant area, which is bounded by $\dot \theta/n \le 1 + \nu/n \approx 1.02$ (Fig.\,\ref{fig:II}).
For any $\tau$ value, the rotation rate of the planet decreases due to tidal effects until it approaches this area, where multiple spin-orbit resonances are present.
In Figure~\ref{fig:V} we show some examples of evolution for different $\tau$ values.

\figV

As in the case of \object{KOI-227}\,$b$, for $\tau \le 10^{-2}$~yr, the spin is in the low frequency regime ($\tau n \ll 1$), and the rotation rate evolves into the pseudo-synchronous equilibrium (Eq.\,(\ref{170207b})).
The average eccentricity of \object{Kepler-88}\,$b$ orbit is 0.065, which gives for the pseudo equilibrium $\langle \dot \theta / n \rangle \approx 1.025$.
The libration width for a maximum value of $C_{22}$ is $\sigma/n \approx 0.008$ (Eq.\,(\ref{170208a})), so in this regime initial prograde rotations never cross any spin-orbit resonances (Fig.\,\ref{fig:V}, top).
However, for $10^{-1} \le \tau \le 10^0$~yr, the equilibrium value is already inside the libration width of the synchronous resonance.
Thus, for these $\tau$ values the spin can be captured in the synchronous resonance (Fig.\,\ref{fig:V}, middle).

For $\tau \ge 10^{1}$~yr the spin is already in the high frequency regime, which means that capture in asynchronous resonances could be possible.
However, the equilibrium $C_{22} \approx 6.7 \times 10^{-6}$ of \object{Kepler-88}\,$b$ is large enough so that the libration zones of individual resonances merge (Eq.\,\ref{170217a}).
As a consequence, as explained in Sect.~\ref{KOI142nrc}, a large chaotic zone around the synchronous resonance is expected.
Indeed, in all simulations we observe a chaotic behaviour for the spin (Fig.\,\ref{fig:V}, bottom).
This result is very interesting as it shows that the rotation of gaseous planets (with a residual $C_{22}=0$) can also be chaotic when its orbit is perturbed by a companion planet.

In Figure~\ref{fig:II} we observe that a small stable synchronous island subsists at the middle of the chaotic zone.
Therefore, we cannot exclude that after some chaotic wobble the spin finds a path into this stable region.
Nevertheless, in our numerical experiments we never observed a simulation where the spin is permanently stabilized in the synchronous resonance.
From time to time the rotation appears to enter the synchronous island, but then the $C_{22}$ grows to a value slightly higher than the theoretical estimation given by expression (\ref{170217a}).
When the maximum deformation is achieved the spin suddenly returns into the chaotic zone.
Indeed, for $C_{22} > 6.7 \times 10^{-6} $ the small resonant island might
totally disappear and the spin might always remain chaotic.

\section{Discussion}
\label{sec:discussion}

We show that close-in planets inside or close to orbital resonances
undergo perturbations of their spins.
For small eccentricities and weak orbital perturbations,
the only spin equilibrium is the synchronous spin-orbit resonance,
for which the rotation period equals the orbital revolution period.
Tidal dissipation naturally drives the spin of the planet into this unique equilibrium,
which is why close-in planets are usually assumed to be tidally synchronised.

When the planet-planet perturbations are significant,
we demonstrate that
new sub-synchronous and super-synchronous spin-orbit resonances appear, even for quasi-circular orbits.
Moreover, for planets observed to transit, the transit timing variations (TTVs)
provide the location (TTV period) and size (TTV amplitude) of these new resonances.
For planets undergoing strong TTVs, the spin could be tidally locked in these asynchronous
states or could even be chaotic.

We apply our modelling to \object{KOI-227}\,$b$ and \object{Kepler-88}\,$b$,
and run numerical simulations of the spin of these planets.
We find that the spin of \object{KOI-227}\,$b$ has a non-negligible probability
of being locked in an asynchronous resonance,
while the spin of \object{Kepler-88}\,$b$ could be chaotic.
In the case of \object{KOI-227}\,$b$, we assume the planet to
be mainly rocky since the bulk density found by \citet{nesvorny_photo_2014}
in the best fitting solution is very high ($18.6\ \mathrm{g\ cm^{-3}}$).
However, the mass of the planet is not well constrained and this planet could have a
non-negligible gaseous envelope.
For \object{Kepler-88}\,$b$ (gaseous planet), in the most realistic cases,
the spin is always locked in the (pseudo-)synchronous state.
To observe a chaotic evolution we need the relaxation timescale ($\tau$, see Sect.~\ref{sec:numsims}) to be at least 10~yr.
This value is very typical for rocky planets, but probably overestimated for a gaseous
planet such as \object{Kepler-88}\,$b$.
Nevertheless, these two cases illustrate very well what
the spin evolution of smaller rocky planets with strong TTVs could be.
We note that the planets with the strongest TTVs found in the literature
\citep{nesvorny_detection_2012,nesvorny_king_2013,nesvorny_photo_2014}
are mostly giant planets, probably due to observational biases.
Indeed, individual transits of small planets are very noisy, and these planets
are usually fitted by phase-folding the light-curve.
This makes small planets undergoing strong TTVs particularly challenging to detect.

For the sake of simplicity of the model we assume coplanar orbits, no obliquity, and low eccentricities for both planets.
However, this is not a limitation for the application to observed systems,
as numerical simulations including these effects can be performed.
In particular, adding some inclination/obliquity will increase the number
of degrees of freedom and probably ease the chaotic behaviour of the spin
\citep{wisdom_chaotic_1984,correia_spin_2015}.

\begin{acknowledgements}
  We thank the anonymous referee for his/her useful comments.
  We acknowledge financial support from SNSF and CIDMA strategic project UID/MAT/04106/2013.
  This work has been carried out in part within the framework of
  the National Centre for Competence in Research PlanetS
  supported by the Swiss National Science Foundation.
\end{acknowledgements}

\bibliographystyle{aa}
\bibliography{biblio}

\appendix

\section{Near-resonance case}
\label{sec:near-resonance-case}

In this appendix we show how the quasi-periodic decomposition of
$\Lambda_i$, $\lambda_i$  (see Eq.~(\ref{eq:Lat}),~(\ref{eq:lat}))
can be obtained in the near-resonance case.
The non-resonant case is the easiest to deal with
since the classical secular approximation can be used.
The perturbative part of the Hamiltonian (see Eq.~(\ref{eq:hamil}))
can be expanded in Fourier series of the mean longitudes
\begin{equation}
  \H_1 = \sum_{l\in\mathbb{Z}^2} C_l(\Lambda_i,D_i,\varpi_i) \expo{\im (l_1\lambda_1+l_2\lambda_2)},
\end{equation}
The secular evolution of the system is obtained by
averaging this Hamiltonian over the fast angles $\lambda_i$ ($i=1,2$).
This averaging transformation is obtained by a change of coordinates that is close to identity.
We denote by $\Lambda_i'$, $\lambda_i'$, $D_i'$, and $\varpi_i'$ ($i=1,2$) the new coordinates,
by $\H'$ the new Hamiltonian,
and by $W$ the generating Hamiltonian of the transformation.
To first order in the planet masses, the change of coordinates reads
\begin{eqnarray}
  \label{eq:chglambda}
  \lambda_i &=& \lambda_i' + \{W,\lambda_i'\} = \lambda_i' + \frac{\partial W}{\partial \Lambda_i'},\\
  \label{eq:chgLambda}
  \Lambda_i &=& \Lambda_i' + \{W,\Lambda_i'\} = \Lambda_i' - \frac{\partial W}{\partial \lambda_i'},\\
  \label{eq:chgw}
  \varpi_i &=& \varpi_i' + \{W,\varpi_i'\} = \varpi_i' + \frac{\partial W}{\partial D_i'},\\
  \label{eq:chgD}
  D_i &=& D_i' + \{W,D_i'\} = D_i' - \frac{\partial W}{\partial \varpi_i'},
\end{eqnarray}
and the Hamiltonian $\H'$ is given by
\begin{eqnarray}
  \H_0' &=& \H_0,\\
  \H_1' &=& \H_1 + \{W,\H_0\},
\end{eqnarray}
with
\begin{equation}
  \H_1' = <\H_1> = C_{0},
\end{equation}
and where the braces denote the Poisson brackets.
We thus have
\begin{equation}
  \label{eq:homological}
  \{W,\H_0\} = <\H_1> - \H_1 = -\sum_{l\in\mathbb{Z}^2\backslash 0} C_l \expo{\im l.\lambda},
\end{equation}
which is called the homological equation and whose solution is
\begin{equation}
  W = \sum_{l\in\mathbb{Z}^2\backslash 0} \frac{C_l}{\im l.n}\expo{\im l.\lambda},
\end{equation}
where $n_i$ ($i=1,2$) are the unperturbed Keplerian mean-motions of the planets.

The secular Hamiltonian $\H'$ no longer depends on the mean longitudes $\lambda_i'$,
which implies that the coordinates $\Lambda_i'$ are constants of motion.
The long-term variations of $D_i'$ (i.e. eccentricities), $\varpi_i'$, and $\lambda_i'$,
can be solved by using the Hamiltonian $\H'$.
However, in this study, we are interested in the evolution of the system at shorter timescales
and will neglect this secular evolution.
We thus assume that $D_i'$ and $\varpi_i'$ are constants and that
$\lambda_i' = \lambda_{i,0}' + n_i t$.
In order to obtain the real evolution of the system, we need to revert to the original coordinate system
using Eqs.~(\ref{eq:chglambda})-~(\ref{eq:chgD}).
In particular, we have
\begin{equation}
  \label{eq:smalldiv}
  \Lambda_i = \Lambda_i' - \sum_{l\in\mathbb{Z}^2\backslash 0}  \frac{l_i C_l}{l.n}\expo{\im l.\lambda'}
\end{equation}
Let us consider a system that is close to a $k_2$:$k_1$ resonance but outside of it.
The combination $\nu = k_2 n_2 - k_1 n_1$ is thus small (but not zero)
which enhances the corresponding terms in
the change of coordinates (small divisor).
Keeping only this enhanced term, we have
\begin{equation}
  \Lambda_i \approx \Lambda_i' + \frac{2 k_i}{\nu}|C_{k}|\cos(\nu t + \phi_i),
\end{equation}
with $k=(-k_1,k_2)$,
$C_{k}=(-1)^{i+1}|C_{k}|\expo{\im \phi_i}$,
and $C_{-k}$ is its complex conjugate (by construction).
The degree of the resonance $k_2$:$k_1$ is denoted $q=k_2-k_1$.
The coefficient $C_{k}$ is of order $q$ in eccentricity (D'Alembert rule),
and can be written \citep[e.g.][]{laskar_stability_1995}
\begin{equation}
  \label{eq:CoefC}
  C_{k} = \frac{\mathcal{G} m_1 m_2}{a_2} e^q c_{k},
\end{equation}
where $e=\max(e_1,e_2)$ and $c_{k}$ is of the order of unity.
Finally, we have (see Eq.~(\ref{eq:defalpha}))
\begin{eqnarray}
  \label{eq:nuquasires}
  \nu &=& k_2 n_2 - k_1 n_1,\\
  \label{eq:alphaquasires}
  \alpha &=& 6 k_i |C_{k}| \frac{n_i}{\Lambda_{i,0} \nu^2}
  \sim \frac{m_p}{m_0} e^q \left(\frac{n_i}{\nu}\right)^2,
\end{eqnarray}
where $m_p$ is the mass of the perturbing planet.
We observe that when the system is very close to the resonance separatrix,
$\nu \ll n$, the amplitude of oscillations increases (see Eq.~(\ref{eq:alphaquasires})).

\section{Resonant case}
\label{sec:resonant-case}

In this appendix we show how the quasi-periodic decomposition of
$\Lambda_i$, $\lambda_i$ (see Eq.~(\ref{eq:Lat}))
can be obtained in the resonant case.
The resonant case arises when the small divisor of Eq.~(\ref{eq:smalldiv}) is
too small and the averaging technique of Appendix~\ref{sec:near-resonance-case}
is no longer valid.
However, the non-resonant terms can still be averaged out using the same procedure
(see Appendix~\ref{sec:near-resonance-case}).
The resonant secular Hamiltonian $\H'$ is constructed such that
resonant terms (of the form $p(k_2\lambda_2-k_1\lambda_1)$) are kept,
\begin{equation}
  \H_1' = \sum_{p\in\mathbb{Z}} C_{(-pk_1,pk_2)} \expo{\im p (k_2\lambda_2'-k_1\lambda_1')}.
\end{equation}
The solution to the homological equation (Eq.~(\ref{eq:homological})) is in this case
\begin{equation}
  W = \sum_{l\in\mathbb{Z}^2, l\neq(-pk_1,pk_2)} \frac{C_l}{\im l.n}\expo{\im l.\lambda},
\end{equation}
and the resonant Hamiltonian reads
\begin{equation}
  \H' = \H_0' + \sum_{p\in\mathbb{Z}} C_{(-pk_1,pk_2)} \expo{\im p (k_2\lambda_2'-k_1\lambda_1')},
\end{equation}
where the coefficients $C_{l}$ are functions of $\Lambda_i'$, $D_i'$, $\varpi_i'$.
In the following we neglect the secular evolution of the eccentricities
and longitudes of periastron as in the non-resonant case
(see Appendix~\ref{sec:near-resonance-case}).
Moreover, the coefficients $C_{l}$ have a weak dependency over $\Lambda'$
\citep[e.g.][]{delisle_resonance_2014},
and in first approximation we assume they are constant.
According to Eq.~(\ref{eq:CoefC}), we have
\begin{equation}
  \label{eq:CoefCb}
  C_{(-pk_1,pk_2)} = \frac{\mathcal{G} m_1 m_2}{a_2} e^{|p|q} c_{(-pk_1,pk_2)},
\end{equation}
where $q = k_2-k_1$ is the degree of the resonance, and
$c_{(-pk_1,pk_2)}$ is of the order of unity.
In a first approximation, we only keep the terms of order $q$ in eccentricity,
and obtain the Hamiltonian
\begin{equation}
  \label{eq:resHam}
  \H' = \H_0' + |C_{-k_1,k_2}| \cos(k_2\lambda_2' - k_1\lambda_1' + \phi),
\end{equation}
where $C_0$ has been dropped since we assumed it to be constant.
We introduce the canonical change of coordinates
\begin{eqnarray}
  \label{eq:psiJ}
  &\psi = k_2\lambda_2' - k_1\lambda_1'+\phi,& J= \frac{\Lambda_{1,0}'-\Lambda_1'}{k_1},\\
  \label{eq:xiGamma}
  &\xi = \lambda_2',& \Gamma = \Lambda_2' + \frac{k_2}{k_1} \Lambda_1'.
\end{eqnarray}
Since $\H'$ only depends on the angle $\psi$ and not on $\xi$,
$\Gamma$ is a conserved quantity.
We now develop the Keplerian part $\H_0'$ in power series of $J$.
Constant terms can be neglected since they do not contribute to the dynamics.
Moreover, the first order terms cancel out for a resonant system.
We thus obtain, up to second order in $J$,
\begin{equation}
  \H_0' \approx - K_2 J^2,
\end{equation}
with
\begin{equation}
  \label{eq:K2}
  K_2 = \frac{3}{2} \left( k_1^2 \frac{n_1}{\Lambda_{1,0}}
  + k_2^2 \frac{n_2}{\Lambda_{2,0}} \right).
\end{equation}
With this approximation, the Hamiltonian~(\ref{eq:resHam}) reads
\begin{equation}
  \label{eq:resHam2}
  \H' = - K_2 J^2 + |C_{(-k_1,k_2)}| \cos\psi,
\end{equation}
which is the Hamiltonian of a simple pendulum.
The frequency of libration is given by
\begin{equation}
  \nu = \sqrt{2 K_2 |C_k|} \frac{\pi}{2} K\left(\sin\left(\frac{\psi_{max}}{2}\right)\right),
\end{equation}
where $K$ is the complete elliptical integral of the first kind,
and the amplitude of libration $\psi_{max}$
is in the range $[0,\pi]$.
For small amplitude oscillations, we have
\begin{eqnarray}
  \nu &=& \sqrt{2 K_2 |C_k|}\nonumber \sim n \sqrt{\frac{m_p}{m_0} e^q},\\
  \label{eq:smallosc}
  \psi &=& \psi_{max} \sin(\nu t + \zeta),\\
  J &=& - \frac{\nu \psi_{max}}{2 K_2} \cos(\nu t+\zeta),\nonumber
\end{eqnarray}
and from Eqs.~(\ref{eq:psiJ}),~(\ref{eq:xiGamma}), we obtain
\begin{eqnarray}
  \label{eq:resLa1}
  \Lambda_1' &=& \Lambda_{1,0}' - k_1 J = \Lambda_{1,0}' + k_1\frac{\nu \psi_{max}}{2 K_2} \cos(\nu t+\zeta),\\
  \label{eq:resLa2}
  \Lambda_2' &=& \Lambda_{2,0}' + k_2 J = \Lambda_{2,0}' - k_2\frac{\nu \psi_{max}}{2 K_2} \cos(\nu t+\zeta).
\end{eqnarray}
In principle, we should revert to the original coordinate system
(coordinates without primes), as in the non-resonant case (see Appendix~\ref{sec:near-resonance-case}).
However, in the resonant case, there are no enhanced terms (with small divisors)
in the change of coordinates $W$,
since we kept these terms in the secular Hamiltonian $\H'$.
We thus have $\Lambda_i\approx\Lambda_i'$.
Finally, from Eqs.~(\ref{eq:defalpha}),~(\ref{eq:resLa1}),~(\ref{eq:resLa2}), we deduce
\begin{eqnarray}
  \alpha &=& \frac{3}{2}\frac{k_i n_i}{K_2 \Lambda_{i,0}}\psi_{max}\nonumber\\
  &=& \frac{1/\Lambda_{i,0}}{k_1/\Lambda_{1,0}
    + k_2/\Lambda_{2,0}} \psi_{max}\nonumber\\
  &\approx& \frac{n_i^{1/3}/m_i}{k_1 n_1^{1/3}/m_1 + k_2 n_2^{1/3}/m_2} \psi_{max}.
\end{eqnarray}
In particular, for $m_1\ll m_2$, we have $\alpha = \psi_{max}/k_1$ for the inner planet (1),
and $\alpha = 0$ for the outer planet (2).
On the contrary, for $m_1 \gg m_2$, we have $\alpha = 0$ for the inner planet,
and $\alpha = \psi_{max}/k_2$ for the outer one.
This is not surprising as the less massive planet undergoes the strongest perturbations.

\section{Permanent and tidally induced deformation}
\label{sec:C22}

In this appendix we show how to estimate the width of the synchronous resonance
($\sigma$) from the known properties of the planets.
From Eq.~(\ref{eq:sigma}), we have
\begin{equation}
  \label{eq:sigmaC22}
  \frac{\sigma}{n} = \sqrt{12 \, \C22i / \zetai} \ .
\end{equation}

The internal structure factor $\zetai$ can be estimated from $\kf$ through the Darwin-Radau equation \citep[e.g.][]{jeffreys_earth_1976}
\begin{equation}
  \zetai = \frac{2}{3}\left(1 - \frac{2}{5}\sqrt{\frac{4-\kf}{1+\kf}}\right) \ .
\end{equation}
We have $\zetai = 2/5$ ($\kf = 3/2$) for a homogenous sphere,
$\zetai\approx 1/3$ for rocky planets,
and $\zetai\approx 1/4$ for gaseous planets.

The deformation of the planet ($\C22i$) has two components,
the permanent deformation ($C_{22,r}$) due to the intrinsic
mass repartition in the planet
and the tidally induced deformation ($C_{22,t}$).
The permanent asymmetry of the mass repartition can be roughly estimated
from the mass and radius of a rocky planet,
and using observations of the solar system planets
\citep[see][]{yoder_astrometric_1995}
\begin{equation}
  \label{eq:C22r}
  C_{22,r} \sim 10^{-6} \left(\frac{R}{R_\oplus}\right)^5 \left(\frac{m}{M_\oplus}\right)^{-5/2}
\end{equation}
For gaseous planets, the permanent asymmetry is very weak ($C_{22,r}\approx 0$).

The tidally induced deformation corresponds
to the adjustment of the planet's mass distribution
to the external gravitational potential.
This deformation is not instantaneous,
and the relaxation time ($\tau$, see Sect.\ref{sec:numsims})
depends on the planet's internal structure.
If the deformation were instantaneous,
the $C_{22,t}$ coefficient would be given
by \citep[Eq.~(\ref{max1}) with $\tau=0$, see also][]{correia_equilibrium_2013}
\begin{equation}
  \label{eq:C22t}
  C_{22,t,\text{inst.}} = \frac{\kf}{4}\frac{m_0}{m}\left(\frac{R}{r}\right)^3\cos 2 (\theta - f),
\end{equation}
where we assume the obliquity to be negligible.
This expression is very similar to Eq.~(\ref{eq:ddtheta}), and using
the same approximations, we obtain an expression very similar to Eq.~(\ref{eq:ddgamma}):
\begin{eqnarray}
  C_{22,t,\text{inst.}} = C_{22,t}^{(0)} \Bigg[\!\!\!\!&&\!\!\!\!\cos 2\gamma\nonumber\\
    \label{eq:C22tdev}
    &&\!\!\!\!+ \sum_j \alpha_j \left(1-\frac{\nu_j}{n}\right) \cos 2\left(\gamma + \frac{\nu_j t + \phi_j}{2}\right)\\
    &&\!\!\!\!+ \sum_j \alpha_j \left(1+\frac{\nu_j}{n}\right) \cos 2\left(\gamma - \frac{\nu_j t + \phi_j + \pi}{2}\right)\Bigg]\nonumber
\end{eqnarray}
with
\begin{equation}
 C_{22,t}^{(0)} = \frac{\kf}{4}\frac{m_0}{m}\left(\frac{R}{a_{0}}\right)^3.
\end{equation}
We assume in the following that the relaxation time $\tau$ is much longer that
the variations of $C_{22,t,\text{inst.}}$,
such that the actual $C_{22,t}$ of the planet is the mean value
of $C_{22,t,\text{inst.}}$ (see Eq.~(\ref{max1})):
\begin{equation}
 C_{22,t} = \left<C_{22,t,\text{inst.}}\right>
\end{equation}
From Eq.~(\ref{eq:C22tdev}),
we deduce that as long as the spin is outside of any resonance,
the tidal deformation average out ($C_{22,t} = 0$).
Therefore, before the resonant capture, the $C_{22}$ coefficient of the planet
reduces to its permanent deformation ($C_{22,r}$).

If the spin is locked in the synchronous resonance,
the averaged tidally induced deformation reaches
\begin{equation}
  \label{eq:C22tsync}
  C_{22,t} = C_{22,t}^{(0)}
\end{equation}
when the amplitude of libration is small ($\sin 2\gamma \approx 0$, $\cos 2\gamma \approx 1$).
Similarly, if the spin is locked in a sub/super-synchronous resonance,
we have
\begin{equation}
  \label{eq:C22tsubsuper}
  C_{22,t} = \alpha_j \left(1\pm\frac{\nu_j}{n}\right) C_{22,t}^{(0)}
\end{equation}
at the centre of the resonance.

Let us apply this reasoning to \object{KOI-227}\,$b$ (rocky planet).
We obtain
$C_{22,r} = 1.4\times 10^{-7}$ for the permanent deformation (see Eq.~(\ref{eq:C22r})).
The tidal deformation is $C_{22,t} = 2.6\times 10^{-6}$ at the centre of the synchronous resonance (see Eq.~(\ref{eq:C22tsync})),
and  $C_{22,t} = 7.9\times 10^{-7}$ at the centre of the sub/super-synchronous resonance
(see Eq.~(\ref{eq:C22tsubsuper})).
The maximum deformation ($C_{22}=C_{22,r} + C_{22,t}$) is thus
$2.8 \times 10^{-6}$ in the synchronous case,
and $9.3\times 10^{-7}$ in the sub/super-synchronous case.

In the case of \object{Kepler-88}\,$b$ (gaseous planet) we obtain
$C_{22,r} = 0$, $C_{22} = C_{22,t} = 2.0\times 10^{-5}$
for the synchronous resonance,
and
$C_{22}=C_{22,t}=5.6\times 10^{-6}$ for the sub/super-synchronous resonances.

We note that all these values assume that the amplitude of libration in the resonance
is vanishing.
We show in Appendix~\ref{sec:forced} that forced oscillations are non-negligible
and significantly reduce the tidal deformation.

\section{Forced oscillations and tidal deformation}
\label{sec:forced}

In this appendix we derive the amplitude of forced oscillations in the
synchronous and sub/super-synchronous resonances,
as well as their implications on the tidal deformation of the planet.
We assume that a single term is dominating the planet's TTVs,
with amplitude $\alpha$,
and frequency $\nu \ll n$.
We neglect here the semi-major axis variations
since their contribution to the spin evolution is of order $\nu/n$
compared to the contribution of the mean longitude variations
(see Eqs.~(\ref{eq:lat}), (\ref{eq:smavar}), and (\ref{eq:ddgamma})).

\subsection{Synchronous case}
\label{sec:forcedsync}

We first assume that the spin of the planet
is locked in the synchronous resonance.
From Eqs.~(\ref{eq:ddtheta3}) and (\ref{eq:lat}) we deduce
\begin{equation}
  \label{eq:ddgammasync}
  \ddot{\gamma} \approx -\frac{\sigma^2}{2} \sin 2\big(
    \gamma
    + \alpha \sin(\nu t +\phi)
    \big),
\end{equation}
where (see Eq.~(\ref{eq:sigmaC22}))
\begin{equation}
  \label{eq:sig2C22}
  \sigma^2 = 12\C22i n^2/\zetai
\end{equation}
and (see Eq.~(\ref{eq:C22t}))
\begin{equation}
 \label{eq:C22forcedsync}
  \C22i
  = C_{22,r} + C_{22,t}^{(0)} \left<\cos 2 \big(
    \gamma
    + \alpha \sin(\nu t +\phi)
    \big)\right>.
\end{equation}
We introduce $h = \gamma + \alpha \sin(\nu t + \phi)$, such that
\begin{eqnarray}
  \label{eq:ddhsync}
  \ddot{h} &=& -\alpha \nu^2 \sin(\nu t + \phi) -\frac{\sigma^2}{2} \sin 2 h,\\
  \label{eq:sig2h}
  \sigma^2 &=& \sigma^2_r + \sigma^2_{t,0} \left<\cos 2 h\right>,
\end{eqnarray}
with
\begin{eqnarray}
  \sigma^2_r &=& 12 C_{22,r} n^2/\zetai,\\
  \sigma^2_{t,0} &=& 12 C_{22,t}^{(0)} n^2/\zetai.
\end{eqnarray}
Equations~(\ref{eq:ddhsync}), (\ref{eq:sig2h}) can be developed
in power series of $h$
\begin{eqnarray}
  \label{eq:ddhsyncdev}
  \ddot{h} &=& -\alpha \nu^2 \sin(\nu t + \phi) - \sigma^2
  \left(h - \frac{2}{3} h^3 + \frac{2}{15} h^5 + ... \right),\\
  \label{eq:sig2hdev}
  \sigma^2 &=& \sigma^2_r + \sigma^2_{t,0} \left<1 - 2h^2 + \frac{2}{3}h^4 + ... \right>.
\end{eqnarray}
At first order (linearized equation), the forced solution is simply
\begin{equation}
  h_\text{lin.} = \frac{\nu^2}{\nu^2-\left(\sigma^2_r+\sigma^2_{t,0}\right)}
  \alpha \sin (\nu t + \phi),
\end{equation}
which is of order $\alpha$.
At higher orders,
odd harmonics of the forced frequency $\nu$ appears,
and $h$ takes the form
\begin{equation}
  \label{eq:devh}
  h = \sum_{k \in 2\mathbb{N}+1} h_k \sin k (\nu t + \phi),
\end{equation}
where $h_k$ is of order $\alpha^k$.
We replace Eq.~(\ref{eq:devh}) in Eqs.~(\ref{eq:ddhsyncdev}) and (\ref{eq:sig2hdev}),
and truncate the resulting expression at a given order $N$ in $\alpha$.
By identifying terms of frequency $k\nu$ ($k \in [1,N]$),
we obtain $N$ polynomial equations on the coefficients $h_1$,...,$h_N$.
Solving this set of equations allows us to determine the coefficients $h_i$,
as well as the corresponding $C_{22}$ value.
The forced oscillations of $\gamma$ are then given by
\begin{equation}
  \label{eq:devgamma}
  \gamma_\text{forced} = \sum_{k \in 2\mathbb{N}+1} \gamma_k \sin k (\nu t + \phi),
\end{equation}
with $\gamma_1 = h_1 - \alpha$, $\gamma_k = h_k$ ($k>1$).

Applying this reasoning to \object{KOI-227}\,$b$, we obtain
$C_{22,t} = 1.8\times 10^{-6}$ (instead of $2.6\times 10^{-6}$)
and $C_{22} = C_{22,t} + C_{22,r} = 1.9\times 10^{-6}$
(instead of $2.8\times 10^{-6}$).
The forced amplitude at frequency $\nu$ is $\gamma_1 = 17^\circ$,
while the amplitudes of harmonics ($3\nu$, etc.) decrease rapidly ($\gamma_{k+2}/\gamma_k \sim 10^{-2}$).
In the case of \object{Kepler-88}\,$b$, we obtain
$C_{22} = C_{22,t} = 6.7\times 10^{-6}$ (instead of $2.0\times 10^{-5}$)
and a forced amplitude of $36^\circ$ (at frequency $\nu$).
As was true for \object{KOI-227}\,$b$, for \object{Kepler-88}\,$b$
the amplitudes of harmonics decrease rapidly.

\subsection{Sub/super-synchronous case}
\label{sec:forcedsubsuper}

We now consider the same effect
(forced oscillations at frequency $\nu$ and its harmonics)
but for the sub/super-synchronous resonances.
We assume that the spin is locked in one of these resonances,
such that
\begin{equation}
  \gamma = \pm\frac{\nu t + \phi_\pm}{2} + \gamma_\text{forced},
\end{equation}
with $\phi_+ = \phi + \pi$, $\phi_- = \phi$.
As for the synchronous resonance, we introduce
$h = \gamma_\text{forced} + \alpha \sin(\nu t + \phi)$ such that
\begin{eqnarray}
  \ddot{h} &=& -\alpha \nu^2 \sin(\nu t + \phi) \mp \frac{\sigma^2}{2}
  \sin (2 h \pm (\nu t + \phi)),\\
  \sigma^2 &=& \sigma^2_r \pm \sigma^2_{t,0} \left<\cos(2 h \pm (\nu t + \phi))\right>.
\end{eqnarray}
As we did for the synchronous case,
these expressions can be developed in power series of h.
Replacing $h$ by
\begin{equation}
  \label{eq:devhsubsup}
  h = \sum_{k \in 2\mathbb{N}+1} h_k \sin k (\nu t + \phi),
\end{equation}
and truncating at a given order $N$ in $\alpha$ ($h_k$ being of order $\alpha^k$),
we obtain a set of $N$ polynomial equations on $h_1$, ..., $h_N$.
We then solve for $h_i$ and determine the corresponding $C_{22}$ value.

In the case of \object{KOI-227}\,$b$, we obtain
$C_{22,t} = 5.2\times 10^{-7}$ (instead of $7.9\times 10^{-7}$) and $C_{22} = 6.6\times 10^{-7}$ (instead of $9.3\times 10^{-7}$),
with a forced amplitude of $\gamma_1 = -6^\circ$ (at frequency $\nu$).
The negative sign means that the forced oscillations
are dephased by an angle $\pi$ with respect to the TTV signal.
In the case of \object{Kepler-88}\,$b$,
we find $C_{22} = C_{22,t} = 2.2\times 10^{-6}$ (instead of $5.6\times 10^{-6}$),
with a forced amplitude of $-10^\circ$.
As in the case of the synchronous resonance,
the amplitudes of harmonics ($3\nu$, etc.) decrease rapidly for both planets ($\gamma_{k+2}/\gamma_k \sim 10^{-3}$).
\end{document}